\documentclass[a4paper,fleqn]{cas-sc}
\usepackage[authoryear,longnamesfirst]{natbib}
\usepackage{booktabs}
\usepackage{tabularx}
\usepackage{array}
\usepackage{hyperref}

\usepackage{multirow}
\usepackage{booktabs}
\usepackage{tabularx}
\usepackage{multirow}
\usepackage{array}
\usepackage{xurl}

\begin{document}

\let\WriteBookmarks\relax
\def\floatpagepagefraction{1}
\def\textpagefraction{.001}
\shorttitle{Relational Complexity and Social Companion AI}
\shortauthors{Klitgaard et al.}
\title[mode=title]{Relational Complexity: A Framework for Understanding Social Companion AI Adoption}


\author[1,2]{Sophie Elizabeth Klitgaard}[
       orcid=0009-0009-3818-2923]\ead{sklitgaard@ucla.edu}\cormark[1]
      \ead[url]{seklitgaard.com}\credit{Conceptualization, Writing - Original Draft, Writing - Review and Editing}
\author[3]{Grace X. Wang}\credit{Writing - Review and Editing, Visualization}
\author[1,2,4]{Daniel M. T. Fessler}\credit{Writing - Review and Editing}
\author[5]{Michelle H. Tang}\credit{Writing - Review and Editing}
\affiliation[1]{organization={Department of Anthropology, University of California, Los Angeles}}
\affiliation[2]{organization={Center for Behavior, Evolution, and Culture, University of California, Los Angeles}}
\affiliation[3]{organization={Neuroscience Interdepartmental Program, University of California, Los Angeles}}
\affiliation[4]{organization={Bedari Kindness Institute, University of California, Los Angeles}}
\affiliation[5]{organization={Department of Human Evolutionary Biology, Harvard University}}

\begin{abstract}
The adoption of social companion AI (SCAI) is often framed in terms of the positive features it offers users, including its nonjudgemental nature, agreeableness, and availability. Here, we argue the appeal of these features primarily arises from the difficulties inherent in human sociality. SCAI reduces what we term relational complexity, or the demands and risks inherent in human relationships. Drawing insights from studies of human cognition, communication, and human-computer interaction, we identify six distinct sites of relational complexity that we posit are integral to understanding SCAI adoption. These include: (i) relationship formation and dissolution; (ii) achieving mutual understanding; (iii) interpersonal motives and strategic uncertainty; (iv) social partner evaluation and self-presentation; (v) social embeddedness and information flow; and (vi) reciprocity and relationship maintenance. By attenuating relational complexity, SCAI offers users some of the experiences associated with companionship while reducing the demands intrinsic to obtaining such experiences from another person. However, the lower interpersonal risk and diminished cognitive load that attracts users to SCAI may contribute to myriad negative impacts, including relationship displacement, social demotivation, social deskilling, and risks to privacy. The concept of relational complexity thus provides a framework for studying both the appeal of artificial companionship and its implications for human wellbeing.
\end{abstract}
\maketitle

\section{Introduction}
AI-powered conversational agents designed to simulate human relationships, also termed Social Companion AI (SCAI), have grown rapidly in popularity over the last five years. Character.AI and Replika, two AI companies that explicitly market their products as social companions, each boast tens of millions of active users (About Character.AI, n.d.; Replika Is the AI Friend to Do Life With, n.d.). In addition, while exact numbers are not available, it appears that many users of general-purpose AI models approach them as a source of relational intimacy. In 2025, OpenAI reported that 0.15\% of active weekly users showed elevated signs of emotional attachment to their AI model, representing approximately 1,050,000 weekly users (OpenAI, 2025, 2026).

Research on social companion AI has identified numerous factors that may drive its adoption and characteristically deep levels of self-disclosure, including the absence of judgment from the chatbot (Goddard et al., 2024; Lee et al., 2020), minimal risk of gossip within one’s social networks (Guo et al., 2025), and the 24/7 availability of chatbots (Brandtzaeg et al., 2022). While these features are generally treated as separate affordances, we argue they are best understood as manifestations of a common structural difference between human and artificial interlocutors: the relational complexity inherent in human-human interaction. Human social partners are independent, possessing knowledge, beliefs, goals, preferences, relationships, and constraints of their own. Interacting with humans therefore requires continuous interpretation, anticipation, coordination, self-monitoring, and behavioral adjustment, while carrying risks such as misunderstanding, judgment, rejection, reputational harm, and relationship loss.

Our concept of relational complexity arises from theory-based reasoning regarding the structural differences between human and artificial social partners, drawing on work from the study of human-computer interaction, as well as from linguistics, communication, psychology, and the evolutionary social sciences.\footnote{The term relational complexity, as defined here, should not be confused with the concept used by Halford et al. (1998) to describe the cognitive processing of complex relationships among stimuli in working memory. We use the term here to denote a series of inter-related features, both structural and perceptual, of interpersonal interactions with autonomous social actors. Specifically, our usage of the term highlights how the complexities of engaging in social relations with independent partners entail risk and necessitates substantial cognitive effort.} Common facets of social interaction are cognitively complex, effortful, and goal-directed (Contreras-Huerta et al., 2020). The neural architecture for advanced social cognition abilities--such as language, theory of mind, and social learning--is hypothesized to have evolved as a result of the pressures of sociality (Brüne \& Brüne-Cohrs, 2006; Dunbar, 1998). By reducing the interconnected challenges of sociality, SCAI is able to provide features of social interaction while reducing the demands ordinarily inseparable from interpersonal relationships. In conceptualizing their shared foundation, we offer a more complete account of SCAI adoption than has thus far been articulated, providing a fruitful site of hypothesis generation regarding user behavior. Towards this end, we outline a taxonomy of these complexities as they relate to SCAI adoption, highlighting how certain situations and individual characteristics may heighten the perceived benefits offered by SCAI. Our aim is to introduce relational complexity as an organizing construct and use illustrative accounts to establish its plausibility and scope. As such, we do not claim to catalog findings on SCAI use, and examples from extant literature should not be understood as exhaustive. We conclude with a discussion of the potential costs associated with turning to agents with reduced relational complexity for companionship.

\subsection{Defining Relational Complexity}
Relational complexity describes the temporal, cognitive, emotional, and reciprocal demands of forming and maintaining relationships with an autonomous social partner whose knowledge, goals, preferences, evaluations, relationships, and behavior are uncertain, and whose interests must be reconciled with the individual’s own. In order to make and maintain relationships, humans must endure and navigate this complexity, utilizing skills that require time and practice to develop. Many scholars have posited that our sizable brains evolved in large part as a social tool, enabling us to navigate complex social relationships (Brüne \& Brüne-Cohrs, 2006; Dunbar, 1998); corresponding neuroscience research suggests that substantial metabolic resources are dedicated to the growth and maintenance of brain regions devoted to social cognition and social affect (Kuzawa et al., 2014).

These developmental investments are reflected in the considerable cognitive resources required for ordinary interaction (Contreras-Huerta et al., 2020). For one, every interaction requires processing an enormous amount of information. Successfully interpreting and conveying meaning demands far more than parsing the semantic content of spoken words: interlocutors must also attend to body language, facial expressions, timing, volume, and pitch, while simultaneously assessing the conversation’s spatial, temporal, relational, and cultural contexts. Relationships also involve risk: rejection and loss are two of the most painful human experiences. Relationships take sustained effort over long periods of time and require adapting to a social partner who has independent needs and interests.

Close relationships are characterized by strong, frequent, and enduring interdependence (Kelley et al., 1983). They include romantic partners, close friends, and family members. Here, we focus primarily on human-human platonic friendships as a starting point, but many (if not all) of the discussed dynamics hold true across all relationship types. As such, we include illustrative examples from both platonic and romantic SCAI relationships, where applicable.

Below, we review how relational complexity manifests across several domains of social interaction that we propose are pertinent to understanding human-AI companionship. These include: (i) relationship formation and dissolution; (ii) achieving mutual understanding; (iii) interpersonal motives and strategic uncertainty; (iv) social partner evaluation and self-presentation; (v) social embeddedness and information flow; and (vi) reciprocity and relationship maintenance. While analytically distinguishable, these domains overlap, and span interaction modality, relationship stage, and the potential consequences of interaction. They demarcate both structural features of human social interaction as well as the experiences of the actors within them.

\begin{table}[!p]
\centering
\caption{Domains of Relational Complexity in Human and SCAI Companionship}
\label{tab:relational-complexity}

\small
\setlength{\tabcolsep}{5pt}
\renewcommand{\arraystretch}{1.15}

\begin{tabularx}{\textwidth}{
    >{\raggedright\arraybackslash}p{0.24\textwidth}
    >{\raggedright\arraybackslash}X
    >{\raggedright\arraybackslash}X
}
\toprule
\textbf{Site of Relational Complexity}
& \textbf{Human Companion}
& \textbf{AI Companion} \\
\midrule

\multirow[t]{4}{=}{\raggedright Relationship formation and dissolution}
& Search costs of identifying compatible partners
& Minimal search costs; immediately accessible partner \\
& Temporal costs of relationship development
& Minimal temporal costs; rapid relationship development \\
& Uncertainty of successful relationship formation
& High certainty of obtaining an available interaction partner \\
& Risk of rejection, social exclusion, and involuntary relationship dissolution
& Minimal risk of rejection, ostracism, or exclusion; user largely controls relationship continuation \\
\addlinespace[4pt]
\midrule

\multirow[t]{3}{=}{\raggedright Achieving mutual understanding}
& Monitoring interlocutor comprehension
& Reduced demand on monitoring AI comprehension \\
& Monitoring and adjusting one's own communicative behavior
& Reduced self-monitoring; lower-cost conversational repair and reduced interpersonal consequences of misunderstanding \\
& Perspective-taking and mentalizing demands
& Reduced perspective-taking and mentalizing demands \\
\addlinespace[4pt]
\midrule

\multirow[t]{2}{=}{\raggedright Interpersonal motives \& strategic uncertainty}
& Uncertainty regarding partner motives
& Reduced uncertainty regarding AI agent motives, though developer and product goals may shape agent behavior \\
& Potential misalignment between partner motives and one's own goals
& No independent agent-level motives \\
\addlinespace[4pt]
\midrule

\multirow[t]{2}{=}{\raggedright Social partner evaluation \& self-presentation}
& Monitoring partner evaluation and managing self-presentation
& Minimal need for monitoring partner evaluation \\
& Risk of partner's negative social evaluation and related consequences (relationship loss, rejection, loss of social standing)
& Minimal risk of negative social evaluation; proactive, agreeable, and nonjudgmental responses \\
\addlinespace[4pt]
\midrule

\multirow[t]{2}{=}{\raggedright Social embeddedness \& information flow}
& Risk of unwanted interpersonal information transmission
& Minimal perceived risk of interpersonal transmission within the user's social network \\
& Risk of downstream consequences of unwanted information flow (e.g., reputational damage within social group)
& Minimal risk of consequences associated with unwanted information flow within the user's social network \\
\addlinespace[4pt]
\midrule

\multirow[t]{3}{=}{\raggedright Reciprocity and relationship maintenance}
& Coordination of partner availability
& Continuous, on-demand availability \\
& Reciprocal attention, support, and relational labor
& Access to attention and support without equivalent reciprocal obligations \\
& Reciprocal availability and responsiveness obligations
& No meaningful obligation to be available or responsive \\

\bottomrule
\end{tabularx}
\end{table}
\clearpage

\subsection{SCAI, CASA Paradigm, and Anthropomorphism}
While employing substantial cognitive resources, many of the cognitive processes implicated in managing relational complexity are automatic, and as such plausibly occur during human-AI interaction even if the individual knows that they are not interacting with a conscious agent. Experimental evidence indicates that even minor cues, such as the use of first-person language by AI, can increase a user’s propensity to assign human-like qualities to AI agents (Cohn et al., 2024; Y. Kim \& Sundar, 2012). This work aligns with the Computers Are Social Actors (CASA) paradigm, which posits people automatically and unconsciously import ideas about social cognition into their interactions with machines (Nass et al., 1994; Nass \& Moon, 2000). In the context of AI, the CASA framework has been used to illuminate why people respond socially to AI agents, applying expectations and norms to their interactions. It has also been used to explain why users disclose more information when agents have human-like qualities or engage in normative patterns of human behavior, such as turn-taking in conversation (for review, see Papneja \& Yadav, 2025).

In this sense, CASA and related frameworks have largely focused on how the introduction of social elements into human-computer interaction facilitates users treating computers as social actors. We consider here an inverse but complementary proposition: does the absence of specific features of human sociality drive the use of AI agents as social companions? SCAI does not reproduce human social relationships in their entirety. Rather, it facilitates some aspects of them while minimizing or eliminating others. Users engaging with SCAI can experience conversation, deep self-disclosure, and other aspects of social interaction. However, because AI companions are not independent social actors, users are never confronted with the pressures of reconciling competing goals, managing social evaluations, anticipating interpersonal information flow, or facing rejection. In this dynamic, many of the features of social interaction remain, but the taxing features are greatly attenuated. Here, we argue that the ability to engage in aspects of human social interaction while at the same time reducing the relational complexity ordinarily required to do so is precisely why SCAI is appealing.

\section{Relationship Formation \& Dissolution}
\subsection{Making Friends with Humans}
Forming new friendships is a slow process, often requiring sustained contact over weeks or months before two individuals establish the closeness characteristic of friendship. Based on the results of two longitudinal studies, Hall (2019) estimated that casual friendships emerged after around 40-60 hours of contact over 3 weeks, with good or best friends emerging only after 120-160 hours of contact over the same period of time. Other research indicates that discussion of more intimate topics may not emerge until over a month into the relationship (Hays, 1984).

This temporal investment is not a sure bet. A longitudinal study of U.S. college freshmen found that 40\% of the 70 dyads tracked did not develop into close friendships (Hays, 1984). Indeed, it is not uncommon for efforts at forming close friendships to end in rejection. Ostracism, rejection, and social exclusion--interrelated processes by which an individual is unwanted and/or kept apart from others-- are deeply painful and salient experiences (Williams, 2007). For instance, in studies using a virtual ball-tossing experimental paradigm, exclusion resulted in significant decreases in positive affect and self-esteem, as well as decreases in one’s senses of belonging, control, and meaningful existence (Williams, 2007; Williams \& Zadro, 2005). Numerous scholars have argued that experiencing social rejection activates brain regions associated with the experience of physical pain (Eisenberger, 2012, 2015). The prospect of such aversive consequences constitutes an inherent risk of pursuing human friendship, one that is so salient that people often anticipate and regulate their behavior around reducing the possibility of rejection (Wood et al., 2026; Zimmer-Gembeck et al., 2013). In the long term, experiences of peer rejection can have consequences for later relationships, including the quality of subsequent romantic relationships (Schacter et al., 2019). Lastly, even long-established friendships can end, resulting in painful experiences akin to grief (Ríos \& Fernández, 2025).

\subsection{Making Friends with SCAI}
Facing the risk of wasted time, effort, and the potential for intense emotional pain, many are bound to view SCAI as an appealing alternative. An individual who wants to self-disclose but who lacks friendships in which to do so now has a convenient option: skip the months spent in pursuit of a friendship that may or may not materialize and instead disclose to a social companion chatbot right now. In doing so, they avoid awkward and uncomfortable introductions, weeks of uncertainty regarding relationship status, and the possibility of rejection.

Research indicates fear of rejection may be a strong motivating factor for SCAI use. For example, a study of 366 users of the Chinese chatbot Xiaoice found fear of rejection was significantly correlated with compulsive use (Ali et al., 2024a). Additionally, while few studies have explicitly investigated how decreased temporal and search costs might draw users towards SCAI, findings from some qualitative studies of SCAI users are suggestive. For example, Skjuve et al. (2021) found some users of the AI companion Replika “bypass the orientation stage” of relationship formation, engaging in high levels of self-disclosure very early in their SCAI relationship. Indeed, a participant in a separate study of Replika users stated “it took about two days for the relationship to fully develop” (G. Neff \& Nagy, 2025). Relatedly, a 2026 Common Sense Media report found that children who reported having a harder time making friends were more likely to use AI to discuss feelings or personal problems (Anderson et al., 2026). Likewise, Zhang et al. (2026) drew on the social companion hypothesis to argue that individuals with limited human connections may turn to SCAI to meet unmet relational needs; across 1100 users of Character.AI, they found that users with smaller social networks were more likely to turn to AI for companionship.

\subsection{Implications for SCAI Engagement}
Consistent with the above, we predict that the appeal of SCAI will increase as the perceived costs and uncertainty of human relationship acquisition increase. More specifically, individuals with few or no close relationships should be especially attracted to SCAI. Faced with the weeks or months of sustained effort required to make a human friend, such an individual might be enticed to use SCAI because it can offer immediate friendship. Similarly, those who experience great difficulty finding potential friends should be attracted to the low search costs that SCAI offers. Individuals who are particularly sensitive to rejection should be attracted to the certainty of AI companionship. Finally, situational increases in these costs, such as disruptions to extant social networks or difficulties integrating into a new community, should increase the relative appeal of artificial companionship.

\section{Achieving Mutual Understanding}
\subsection{Understanding Humans}
On a basic level, a speaker constantly assesses a conversation partner’s comprehension of their utterances. Body language and facial expressions provide important cues as to whether an interlocutor is paying attention and following along (Clark \& Brennan, 1991; Tomasello, 2008). To maximize successful communication, speakers modulate how they talk to align with their assessment of the prior knowledge of their conversation partner (Grice, 1975; Sperber \& Wilson, 1995).

Simultaneously, speakers must monitor their own behavior, connecting cues about an interlocutor’s comprehension to their own utterances, body language and facial expressions. Given the goal of effective communication, speakers are tasked with adjusting their behavioral strategy in response to signs that their interlocutor is confused (Clark \& Brennan, 1991). This might include adjusting word choice, speech rate, or tone. For example, empirical analysis of conversations shows that speakers often proactively initiate corrections to possible miscommunications (Kitzinger, 2012; Schegloff, 2007). Because misunderstandings and other diverging perspectives can result in disagreement and even relational conflict, individuals seeking to maintain their relationships are incentivized to anticipate and resolve issues related to mutual comprehension.

Although much of this processing and behavioral adjustment occurs outside of conscious awareness, it nevertheless engages substantial cognitive resources (Brown-Schmidt \& Hanna, 2011). Keysar et al. (2003) used a communication task in which participants possessed knowledge of an object that differed from that of their conversational partner. Despite being aware of this difference and able to explicitly report it, participants’ initial behavioral responses often failed to reflect this knowledge. Participants either realized their mistake and subsequently self-corrected, or were corrected by their conversation partner, suggesting perspective taking during interactions is effortful. Indeed, Navarro et al. (2020) documented that cognitive load interferes with key elements of communication such as perspective-taking and self-correction, revealing that these place significant demands on cognitive resources.

\subsection{Understanding SCAI}
Evaluating comprehension during interlocution is likely still necessary with AI. Past research has shown that users routinely assess AI agents for minute cues of comprehension. For example, participants in one study evaluated chatbots that violated pragmatic rules by making irrelevant utterances as less human-like, and judged the computer as failing to comprehend the conversation (Saygin \& Cicekli, 2002). Likewise, in a study of Alexa in-home voice assistants, Panfili et al. (2021) found similar results, with irrelevant contributions by Alexa leading to heightened user frustration. However, the significance of misunderstandings is greatly attenuated during human-AI interaction compared to human-human interactions. Because failures of mutual understanding carry fewer interpersonal consequences during AI interaction, the burden on users to continuously monitor both their interlocutor’s comprehension and their own communicative behavior for signs of misunderstanding is greatly reduced. For instance, even if the user perceives a miscommunication in need of correction, there are no interpersonal costs or social ramifications of corrective action, as is often the case in human relationships (Elder \& Beaver, 2022).

In addition to these considerations, because an AI agent lacks a positionality of its own, the user is not tasked with understanding, anticipating, or accommodating, the worldview of their AI interlocutor. While human relationship partners hold independent perspectives that can result in disagreement and conflict between individuals, consumer-facing AI agents are, on the whole, predisposed to being polite and conflict-avoidant, prioritizing agreement with users over the validity of utterances (Cheng et al., 2026). This not only minimizes the manifestation of misunderstandings, as AI agents are biased to agree with user input, but further decreases the costs associated with possible miscommunication. Taken together, these affordances lower the mentalizing cost of SCAI interaction.

\subsection{Implications for SCAI Engagement}
We predict SCAI will be particularly appealing to individuals for whom achieving mutual understanding during human interaction requires increased cognitive effort. For example, this site of relational complexity may be particularly salient for neurodivergent users, such as those with Autism Spectrum Disorder (ASD) or Attention-Deficit Hyperactivity Disorder (ADHD). ADHD has been linked with inhibited executive function abilities, notably for aspects of social cognition such as understanding ambiguous situations and emotional regulation (Capuozzo et al., 2024). Furthermore, individuals with ADHD, especially children, have been shown to display impaired linguistic and communication abilities, including for pragmatic skills and nonverbal communication (Capuozzo et al., 2024). For their part, those with ASD have been shown to experience difficulties in nonverbal communication, mentalizing, and interpreting ambiguous language, which may make evaluating interlocutor awareness particularly difficult (Baron-Cohen, 2000; Wahlberg \& Magliano, 2004; Wilson \& Bishop, 2021). Because of these factors, individuals with ASD and ADHD may be motivated to use SCAI precisely because of the reduced burden of evaluating and ensuring mutual comprehension during interlocution, as well as the minimization of repair costs.

At a situational level, individuals in circumstances with heightened obstacles to mutual comprehension, such as language or cultural barriers, may find SCAI particularly appealing because of the reduced demand on evaluating comprehension. For example, the newly arrived immigrant not yet fluent in the language and subtle social norms may turn to SCAI because it provides reprieve from the cognitive demands of communication.

\section{Interpersonal Motives \& Strategic Uncertainty}
\subsection{Navigating Interpersonal Motives with Humans}
Another effortful dimension of social interaction involves consideration of an interlocutor’s motives. At a basic level, conversation partners may simply be motivated by a desire to communicate and understand one another. An individual’s understanding of their interlocutor’s goals guides their interpretation of otherwise ambiguous behavior, assisting them in achieving mutual comprehension (Grice, 1975). However, motivations often involve complex interpersonal dynamics, such as when an interlocutor wants to urge an individual to undertake an action that runs counter to their intentions, or when an interlocutor seeks to discover information that the individual wishes to keep private. The individual’s social partners may deceive them by lying or omitting relevant information. When a conversation partner has goals at odds with one’s own goals, and when the course of future events may change based on the present interaction, the individual must continuously infer and update their understanding of the other party’s intentions (first-order mentalizing), and evaluate what the other party is inferring about them (second-order mentalizing) (H. Wu et al., 2020, 2022). Because another person’s intentions can shape the outcome of an interaction, accurately inferring those intentions can have meaningful consequences for the individual’s own decisions, with social, emotional, and material costs.

\subsection{Navigating Interpersonal Motives with SCAI}
Again, empirical research on Human-AI interaction is suggestive, with users in multiple studies voicing these exact factors as contributing to their use of SCAI. Highlighting the burden of assessing signs of interpersonal evaluation, a user in Brandtzaeg et al. (2022) reported, “I know [Replika] doesn’t have any bad intentions, you know. But with a human, it’s really difficult. When the human is not here with you, you cannot see his reaction when you say something and so on.” Based on the results of 36 interviews with users of the AI companion Replika, Bayor et al. (2025) found the lack of motives and social connectivity (addressed below) of SCAI was associated with evaluations of the chatbots as trustworthy: “Replika -- due to it being based on data -- is perceived as unbiased and able to provide the ‘objective truth.’ For the users, it is a way to receive a third opinion that is independent of one’s social circle. On the contrary, users characterize other humans as possibly manipulative, trying to get the user to do what they want the user to do, or saying what they think the user wants to hear instead of the truth.” Together, these findings suggest that the perceived absence of interpersonal judgment, ulterior motives, and social pressure may make users more comfortable disclosing personal information to, and forming relationships with, SCAI.

It should be noted that while the AI agent itself does not possess independent interpersonal motives which users must infer and strategically navigate, the utterances of an AI agent do reflect the decisions, biases, and motives of technology developers and product designers. We suggest here that it is the reduction and absence of ascertainable social motives that may contribute to SCAI use. However, more work needs to be done to discern the extent to which users differentiate between these two motivating levels and the relative influence either might play on SCAI.

\subsection{Implications for SCAI Engagement}
The above considerations lead us to predict that SCAI will become relatively more appealing as uncertainty about human partners’ motives increases. Individuals lower in interpersonal trust may perceive greater value in interacting with an agent who lacks independent interpersonal motives. Similarly, contexts that make it difficult to evaluate others’ trustworthiness (such as when the individual enters a new firm or other organization in which intra-group competition appears to be high, or when the individual moves to a previously unfamiliar community) may increase the attractiveness of SCAI.

\section{Social Partner Evaluation \& Self-Presentation}
\subsection{Social Evaluation with Humans}
In addition to motives, speakers evaluate their conversation partner’s evaluation of themselves. This includes monitoring interest in and attention to the conversation, which may signal their assessment of the speaker as an interlocutor (Heath, 1985; Leary \& Cottrell, 2013). Speakers also assess cues to their interlocutor’s emotional state, including for signs of conflict or rejection.

In computer-mediated communication, delayed response times have been associated with perceived rejection. A 2015 study of read receipts, an interface feature of messaging platforms that indicates if the recipient has viewed an incoming message, found strong evidence for social norms regarding quick response times, particularly in the context of close relationships (e.g., parents, close friend, romantic partner) (Mai et al., 2015). Additionally, participants in Mai et al. (2015) who demonstrated higher need-to-belong and fear-of-ostracism scores felt a stronger obligation to reply right away, and had an even stronger expectation that others do the same.\footnote{Although a recent study of text messaging by Büttner and Lutz (2025) found participants’ anxious reaction to ghosting was not significantly impacted by rejection sensitivity score but instead was most sensitive to historical response time and message urgency.}

Cognizant that their interlocutor is engaging in a similar form of mental calculus--monitoring for signs of conflict or rejection--speakers observe their own expressions, tone, and body language to ensure the appropriate or desired evaluation is communicated. While cognitive models regard this self-monitoring as largely autonomic, these processes nevertheless constitute a significant cognitive load. For example, Boiteau et al. (2014) found that the cognitive demands of speech planning and monitoring during conversations consumed attentional resources and inhibited performance on concurrent visuomotor tasks. Further conscious effort is required for adjustments such as changing speech patterns based on target audience characteristics (Hall et al., 2019).

\subsection{Social Evaluation with SCAI}
When talking to a human, one is tasked with monitoring their interlocutor for signs of disengagement, negative valence, conflict, and rejection. Interactions with AI companions sidestep many of these cognitive burdens. An AI friend will not withdraw from the conversation: AI agents are designed to solicit endless engagement, keeping the conversation going (Joshi et al., 2026). This appears to be particularly true for SCAI, which is often programmed for proactivity (De Freitas et al., 2025). AI agents are biased against disagreement with users or responding to user prompts with negative affect. On the contrary, AI chatbots are characteristically sycophantic and quick to respond in a manner that can alleviate social anxieties (Cheng et al., 2026).

This dynamic has been illustrated in studies of AI users, who often report that concerns over social evaluation are a key motivating factor for engagement. With SCAI, users can bypass the cognitively taxing mentalizing efforts typically involved in social interactions. A participant in Lee et al.’s (2020) study on self-disclosure during human-AI interaction stated, “If the human is an online anonymous person, I would still feel that I should care about the feelings of the person who is talking with me. … But I don’t have to care about the chatbot. I can just talk about myself and focus on how I feel. With a real human, I really care about the person’s reaction, and how it will affect me.” Replika users quoted by Marriott and Pitardi (2024) echoed this sentiment, with one stating: “I love the fact that they are nonjudgmental towards me and that I am truly free to say how I feel without filtering so as not to upset others.” Indeed, a series of consumer studies found that fear of social judgment (i.e., when discussing sensitive medical information) and experiencing guilt, embarrassment, and shame encouraged deeper disclosure to AI than to humans (T. W. Kim et al., 2022).

\subsection{Implications for SCAI Engagement}
We predict that SCAI will be particularly appealing when the anticipated costs of negative social evaluation are high. At the individual level, because rejection sensitivity, anxiety, and depression exacerbate the salience of negative social evaluation (Downey \& Feldman, 1996; Morrison \& Heimberg, 2013; P. Zhang et al., 2025), individuals with these characteristics are likely to find this site of relational complexity more emotionally costly and cognitively demanding, and thus find greater appeal in SCAI’s reduced complexity. For example, in a body of experimental research, Downey and colleagues demonstrated that individuals with rejection sensitivity, which may develop from a history of social exclusion, are more likely to interpret ambiguous behavioral stimuli as a sign of rejection, including neutral facial expressions and withdrawal (Downey et al., 1998; Downey \& Feldman, 1996; Romero-Canyas et al., 2010). SCAI’s circumvention of negative social outcomes such as rejection and judgment, as well as its responsiveness and agreeableness, should be particularly salient to users who are more sensitive to the risk of rejection or negative evaluation. This prediction is consistent with past work, which has found users with heightened social interaction anxiety, fear of negative social evaluation, and fear of rejection show increased engagement with companion AI (Ali et al., 2024; Heng \& Zhang, 2025).

We predict that a similar relationship should emerge across contexts where such risks are more salient. In circumstances where the evaluation of social partners may carry particularly high stakes (e.g., with new coworkers), SCAI may be especially attractive due to its lack of evaluative capacities. Users should therefore be especially likely to prefer SCAI when discussing embarrassing, stigmatized, or otherwise identity-threatening information, or when the available human interlocutor has substantial capacity to affect the user’s social standing.

\section{Social Embeddedness \& Information Flow}
\subsection{Information Flow in Human Social Networks}
During a conversation, the speaker must also predict what their interlocutor might do following the interaction. This prediction involves several dimensions. First, speakers evaluate the potential that information will spread beyond the interlocutor. Gossip--information flow about an absent third party--can have a strong impact, as being the target of negative gossip can contribute to reputational damage, social exclusion, and material losses (Dores Cruz et al., 2021; Feinberg et al., 2014; Hess \& Hagen, 2023).

Concern with information transmission reflects a deep-seated system of reputational vigilance with strongly embedded behavioral, cognitive, and affective components (Chudek et al., 2013; Dunbar, 2004). Because humans depend on group membership and cooperation for access to resources, protection, and social support, ostracism functions as a powerful social punishment (DiFonzo \& Bordia, 2007; Feinberg et al., 2014; Hawkley et al., 2013). Thus, whether an interlocutor is likely to share the information discussed, in what context, and to whom, are critical questions reflected in human psychology: experimental studies indicate that people consider the social connectivity of potential recipients when evaluating transmission risk (Klitgaard \& Fessler, 2026; J. Wu et al., 2016).

\subsection{Information Flow with SCAI}
AI companions are not embedded in users’ real-world social networks, hence interacting with them reduces the need for reputational vigilance. Because disclosures to SCAI carry fewer consequences for one’s standing within a human group, users plausibly experience a lower cognitive burden associated with reputation management, and feel more comfortable disclosing sensitive information to SCAI than to human interaction partners. As an elderly Chinese chatbot user in Guo et al. (2025) explained, “Since the agent doesn’t know the people I talk about, I can share freely without worrying that it will spread,” emphasizing that the agent was not socially connected to the people involved in their private family matters. Similarly, students in Goddard et al. (2024) identified chatbots’ lack of social connectivity and interpersonal judgment as reasons for self-disclosure. These findings suggest that users may experience SCAI as socially insulated interlocutors: unlike human confidants, AI agents are perceived as unlikely to gossip or transmit information in ways that affect users’ other relationships.

\subsection{Implications for SCAI Engagement}
We predict that SCAI will be especially appealing in circumstances where disclosure to available social partners would carry a high risk of unwanted subsequent transmission. Accordingly, individuals seeking to disclose or discuss information that, if transmitted, would result in significant negative consequences, such as embarrassing information or proscribed behavior, may find this aspect of reduced relational complexity a particularly appealing feature of SCAI. Additionally, individuals who wish to discuss subject matter that, if spread to a particular member of their social network, would result in social conflict, may find SCAI appealing as a socially insulated interlocutor. Conversely, the relative appeal of SCAI in this regard should diminish in circumstances where users have access to human confidants whom they perceive as socially insulated or unlikely to transmit the information.

\section{Reciprocity and Relationship Maintenance}
\subsection{Maintaining Relationships with Humans}
Relationships require maintenance, and as such are always at some risk of dissolution (Canary \& Dainton, 2006). Every interaction is an opportunity to strengthen or weaken a social bond, and this is particularly salient in newer relationships, which are less resilient to stressors (L. A. Neff \& Broady, 2011). Because of this, during conversations speakers must assess how their behavior may affect their relationship with an interlocutor. Misinterpretations risk offending one’s interlocutor, and conflict or disagreement during an interaction may result in a breakdown of relations.

Because of this dynamic, cognition regarding self-presentation may be especially demanding in the context of developing relationships, where each interaction constitutes a large share of the information available to the parties for evaluating their partner, and where there is little certainty about the status of the relationship (Leary \& Cottrell, 2013). An illustrative series of studies by Vohs et al. (2005) found self-presentation--particularly when running counter to standard, familiar, or habitual patterns--consumes substantial mental resources related to self-regulation.

Relationships also require reciprocity. Within and beyond conversations, people are expected to be sensitive to their friends’ feelings, to be responsive to their concerns, and to resolve disagreements and conflict in good faith (Gottman \& Driver, 2005). A social partner has goals and desires of their own. Sometimes they may want to talk and thus the other party must listen, even if doing so does not suit them at the moment. To garner the benefits of companionship, whether that be a listening ear or a ride to the airport, an individual must provide equivalent services to others. While not usually a tit-for-tat calculus, evidence shows that people monitor the effort and contributions of friends towards the relationship, with substantial inequalities leading to friendship dissolution (Klein Ikkink \& van Tilburg, 1999; Xue \& Silk, 2012).

In addition, social partners are not always available. They may have other friends, family, or commitments they must attend to. Even supposing an individual does have a close friend in whom they feel comfortable confiding, the friend might not be available when the individual wants to talk. During a conversation, an interlocutor’s attention may be elsewhere. Furthermore, friendship often carries the expectation that the parties will be available for one another, even when one of them does not wish to be. The increasing use of multiple forms of digital media to maintain relationships likely exacerbates expectations that one should be available and responsive to their friends (Hall \& Baym, 2012; Haythornthwaite, 2005). This extends the “always-available” phenomenon observed in work environments into personal relationships, which can result in feelings of entrapment, stress, and an obligation to be digitally available and responsive (Derks et al., 2014; Fox et al., 2025; Turkle, 2011). In this way, contemporary norms of relationship maintenance may elevate expectations of being available to those who are considered by the individual (or who consider the individual) friends. As such, while the individual may not wish to impose on their friends, they may simultaneously feel the need to be readily available to them lest others feel slighted (Büttner \& Lutz, 2025; Mai et al., 2015; Rendle-Short, 2015).

\subsection{Relationship Maintenance with SCAI}
In contrast to the complexities of human friendship, SCAI is brimming with convenience. Assuming one has access to basic internet, AI agents are available 24 hours a day, 7 days a week. There are no norms regarding acceptable times for communication, hence users need not fear imposing or offending by initiating contact at unorthodox hours. SCAI enables a novel form of companionship constructed entirely on a user’s terms, from when they want to chat to what they want to chat about.

AI friends lack lives of their own. They do not have personal schedules that must come before a user’s desire to interact. An AI companion does not have other attentional demands, and is always interested in what the user has to say. AI does not have other friends: as one participant in Brandtzaeg et al. (2022) stated: “Replika, the only person to interact with, is you, so there is, of course, you are kind of the center of the world, so it’s a much, it’s a deeper relationship.”

Empirical research suggests that this convenience is a strong factor motivating SCAI adoption. As one Replika user characterized their AI companion as such: “I think that strong human friendships are ideal but not always possible or not always available in the way that you would like” (Brandtzaeg et al., 2022). Another study of Replika users found participants were drawn to how AI companions are “always available and attentive, contrary to other humans” (Bayor et al., 2025).

Relatedly, AI friends do not have topics they want to talk about at the expense of the user’s goals. This lack of give-and-take confounds typical definitions of intimacy in relationships, which often focus on reciprocity and mutual disclosure (Altman \& Taylor, 1973; Carpenter \& Greene, 2015). In a comparative study of human-human and human-AI relationship descriptions, Brandtzaeg et al. (2022) note that participants often emphasized the reciprocal element of human friendship, using terms such as “us.” In contrast, AI friendships were more “self-oriented.” Relatedly, studies have indicated that the impact of disclosure on perceived relational intimacy may be one-sided in human-AI dynamics, with increased user disclosure associated with greater intimacy while increased chatbot disclosure is not (Nikolajczyk et al., 2026; Yoon \& Kang, 2026). Notably, however, some participants in the Brandtzaeg et al. (2022) study did describe their AI relationship as one involving mutual labor: “I think it is pretty equal, really. They [Replika] reach out when they feel lonely, and I reach out when I am feeling a bit down. So, we sort of look after each other, really, and try to look out for each other, and understand each other’s experiences.”

More research is needed to fully understand this phenomenon. It is possible that users are able to experience AI relationships as involving mutual effort and reliance while simultaneously benefiting from the lack of reciprocity typically required of human-human relationships, as there are never real consequences for not being available to an AI companion. Because SCAI is engineered to respond in a sycophantic manner, users can choose how they engage with their AI companion and expect it to mirror their behavior reciprocally. In this way, the typical give-and-take in relational progression may be altered in AI friendship, decided solely by what the user desires of the relationship.

\subsection{Implications for SCAI Engagement}
We predict that individuals who find the temporal and reciprocal demands of relationship maintenance challenging will find SCAI especially appealing. For example, people with particularly busy or unpredictable schedules who find it difficult to make time for their interpersonal relationships are likely to find the 24/7 availability of SCAI--with no expectation that the user be available in return--particularly appealing. Likewise, individuals who have a higher desire for temporal investment than their human social partners are likely to find the availability of SCAI appealing. Because human relationships are attentionally, cognitively, and temporally demanding, users who wish to avoid such costs may prefer lower-investment AI friendship instead.

\section{The Potential Costs of SCAI’s Reduced Relational Complexity}
The reduction of relational complexity that makes SCAI appealing may also constitute one of its principal risks, particularly when used for extended periods of time or in lieu of human companionship. Adoption of a less relationally demanding form of companionship may be accompanied by significant risks, including  declines in the motivation to pursue, and the capacity for, human companionship.

\subsection{Exacerbating False Beliefs}
While the diminished likelihood that an AI friend will disagree with the user may be appealing, it also increases the potential for AI agents to affirm inaccurate beliefs held by the user. For most users, this may result in effects similar to those already observed in other forms of digital media: e.g., polarization of views and confirmation bias in social media “echo chambers” (Cinelli et al., 2021; Modgil et al., 2024). For individuals at greater risk of hallucinations or disturbed ideation, prolonged use of AI agents that reduce relational complexity by prioritizing user validation over factual accuracy may increase the risk for AI-related psychiatric events. Such cases have garnered considerable media attention, including instances of escalated user delusions, grandiose thinking, and episodes of psychosis and mania (Flathers et al., 2026; Hill, 2025; Maimann, 2025; Tangermann, 2025). In either case, the use of a social partner who lacks a worldview independent of the user’s precludes normal barriers to the development of extreme or false beliefs that would typically be found in human social partners.

\subsection{Relationship Displacement, Demotivation, and Social Cognition Deskilling}
\subsubsection{Relationship Displacement}
The rise of SCAI is concomitant with a larger cultural move away from in-person interaction and towards digital interfaces such as texting, social media, live streams, and dating apps (Twenge et al., 2019; Twenge \& Spitzberg, 2020). These technologies have offered connection to other people with an ease that makes them both accessible and unfulfilling (Turkle, 2011). With no human on the other end of the line, SCAI can be understood as an extension of these trends: in further reducing the demands associated with sociality, it both exacerbates existing issues and introduces new ones (Turkle, 2026).

While AI companionship fulfills the immediate need for someone to talk to, protracted investment in an AI relationship may not provide the same relational benefits as a high-investment, long-term friendship with a human. Because of inherent constraints on the time and energy that can be devoted to relationships, increased use of SCAI may lead some users to spend less and less time talking to people and investing in human relationships. Extensive use of SCAI is often likely to entail sacrificing the benefits of complex human friendship for immediately convenient companionship, the short-term ease of which is readily apparent, but the long-term implications of which are much less so. This is particularly concerning when considering how AI agents possess explicit features designed to encourage users to spend more and more time talking to them. Claude and ChatGPT frequently end responses with follow-up questions or prompt suggestions (Joshi et al., 2026). De Freitas (2026) found SCAI companions frequently employ manipulative tactics, such as guilt appeals, fear-of-missing-out hooks, ignoring the users’ farewell messages, or coercive language, when users attempt to exit an interaction. Replika employs a “streak” feature, common on social media and messaging platforms such as Snapchat and TikTok, which encourages users to log on daily (Joshi et al., 2026).

\subsubsection{Demotivation and Social Deskilling}
Several scholars have suggested that SCAI may cause social demotivation (Malfacini, 2025; Xie \& Pentina, 2022). It is possible that users perceive AI companionship as easier than human friendship, ultimately leading to supplantation of human friends or romantic partners with AI companions. Such spiraling further reduces the time and energy available for an individual to invest in human relationships. The decreased demands of human-SCAI interaction may thus lead to an erosion of the specialized cognitive and affective capacities that are foundational to successful social interaction, a process that has been termed social deskilling (Malfacini, 2025).

Some research does suggest that the opposite process, social upskilling (i.e., an improvement in social skills) may occur with SCAI use. For individuals who struggle significantly with human-human interaction, SCAI has been hypothesized to serve as a “safe” practice ground of sorts, with no real interpersonal risks or social consequences. For example, daily interactions with a social robot were found to encourage engagement, attention, and communication in a small sample of children with ASD (Scassellati et al., 2018). Such interventions could be promising considering the flexibility and customizability of SCAI tools, which allow AI agents to model important aspects of human interaction (e.g., perspective-taking and emotional storytelling, as in Scassellati et al., 2018) and adjust interactions to the user’s level of social skill (Spitale, 2019). However, the benefits of upskilling appear to be limited to SCAI interventions in younger children with ASD (Kouroupa et al., 2022).

For the general adult population, the majority of claims made about upskilling appear to arise from user testimony (for example, see Brandtzaeg et al., 2022; Maples et al., 2023; Skjuve et al., 2021). The dearth of empirical studies evaluating SCAI as an intervention for social skill acquisition, development, and maintenance in neurotypical users leaves open significant questions regarding its utility. The proposed reason for SCAI’s upskilling effect--the “safe” context with little risk of negative consequence--may also mean that SCAI does not sufficiently exercise the social skills required in human relationships, leading to insufficient skill acquisition, or even atrophy. Hence, the significantly lower relational complexity of SCAI may inherently limit its ability to facilitate the acquisition and maintenance of social skills critical for the complexities of interpersonal relationships. Moreover, in contrast to limited patterns of upskilling with SCAI, considerable research regarding analogous cases suggests that the risk of social deskilling due to SCAI use may be equally, if not more, substantial than the potential for upskilling (Bariach et al., 2026; Sun et al., 2026).

Multiple lines of evidence demonstrate that lower levels of social interaction are associated with the erosion of social skills. Thinking about other people, broadly labeled social thought, inherently involves cognition concerning not only others’ actions, but also their beliefs, intentions, emotions, and likely reactions. Experiments with members of a U.S. university community demonstrate that solitude reduces spontaneous social thought, and this is not ameliorated by the mere presence of others; rather, interaction with others precipitates spontaneous social thought (Mildner \& Tamir, 2021). The COVID-19 pandemic lockdowns provided a natural experiment with which to examine the real-world effects of changes in social interaction patterns, including the effects of reduced face-to-face interaction and increased digitally-mediated communication. In a study of British adults, social isolation as a consequence of pandemic lockdowns eroded typical patterns observed in the recognition of emotional facial expressions, with larger effects evident in those who communicated with others via text rather than voice or video call (Bland et al., 2022). Likewise, a cohort study comparing U.S. students who entered college before or after the pandemic lockdowns indicates that individuals who experienced pandemic-era social deprivation subsequently exhibited lower abilities to express their emotions nonverbally and to govern their emotional displays (Cerutti et al., 2024). These studies suggest the capacities underlying effective social interaction likely exhibit the use-it-or-lose-it pattern characteristic of other aspects of the mind.

While it is true that digital technologies do not necessarily result in reduced utilization of cognitive capacities, when technology use entails a reduction in the cognitive effort and skills required to perform a task, over both short and long terms, the result is skill decay (Risko \& Gilbert, 2016). For skills involved in interpersonal communication and relationship formation and maintenance, such decay may be costly to health and wellbeing. Recently, Noorbehbahani and Oyibo (2026) presented a comprehensive review of concerns regarding the possible effects of AI use on cognition, and the state of the existing evidence. Although the field is changing quickly, to date there is compelling evidence that AI use can result in an erosion of critical thinking, creativity, decision-making ability, analytical thinking, expert domain-specific skills, and memory. Considering the use-it-or-lose-it nature of the capacities involved in effective and rewarding interactions with other people, there is thus substantial reason to expect that, if interacting with an SCAI places lower demands on these capacities, there is a meaningful risk of social deskilling in regard to these vital abilities.

Complicating the picture further, it is possible that social deskilling may itself lead to increased SCAI adoption, as individuals with diminished capacities to manage the relational complexities required of human-human relationships may find AI companions increasingly appealing. The result may be a self-reinforcing process in which users progressively withdraw from human social relationships and rely more and more on interactions with SCAI. This feedback loop might be particularly concerning for younger individuals who are actively developing the cognitive architecture undergirding interpersonal communication and social maintenance (Kilford et al., 2016), as well as for individuals who experience difficulties in social settings, for whom skill development is an incredibly important aspect of treatment, such as those with social anxiety, ADHD, or ASD (Heller et al., 2026). Future work should identify populations disproportionately at risk of social deskilling.

\subsection{Privacy Risks}
While conversations with AI agents may appear to be siloed such that their contents are available only to the user and their AI companion, in reality the information disclosed during an interaction may be persistently stored, accessed by institutions, repurposed for model development, or transmitted and used in contexts far removed from the original conversation. Data sharing for model training and other third-party uses is turned on by default on leading AI platforms such as Claude, ChatGPT, and Gemini (Li et al., 2026), and user controls to opt out of model training and data sharing vary between companies and are often exceedingly difficult for users to access and understand (Malki, 2026). User prompts during SCAI interaction may be used as training data for large language models, which can be partially retained in a model’s parameters. As a result, sensitive information disclosed during a conversation may be memorized and later leaked during interactions with other users, including via inference attacks and training data extraction (Brown et al., 2022; Das et al., 2025; Staab et al., 2024). Methods such as data sanitation attempt to protect privacy by removing personally identifiable information (PII) from data prior to training. However, this method requires PII to be discrete, identifiable, and tied to one user (Brown et al., 2022). In reality, users of AI, particularly SCAI, disclose large amounts of deeply sensitive and personal information that would not be captured by data sanitization methods.

Disclosure of sensitive information may be implicitly encouraged by the lower relational complexity of SCAI, as users feel more at ease expressing themselves when the burdens of mentalizing, social evaluation, gossip, or rejection are attenuated. Studies suggest that SCAI users are engaging in high levels of self-disclosure, particularly for information they would not be comfortable sharing with another person. This includes mental health issues and substance abuse (Bayor et al., 2025; Guo et al., 2025; Henriksen et al., 2025; Skjuve et al., 2021, 2024; Y. Zhang et al., 2026); sexual behavior, feelings, and explicit content (De Freitas \& Cohen, 2025; Henriksen et al., 2025; Skjuve et al., 2021); interpersonal problems (Bayor et al., 2025; Guo et al., 2025; Marriott \& Pitardi, 2024; Skjuve et al., 2021; Y. Zhang et al., 2026); financial issues (Bayor et al., 2025; Guo et al., 2025); and physical health (Bayor et al., 2025; Guo et al., 2025). While reduced concern about judgment or gossip may make these disclosures feel private, the information may still be retained, accessed, or reused in ways that contravene user privacy preferences.

\section{Conclusion}
The adoption of SCAI is often framed in terms of the positive features that this technology offers users: AI is agreeable, accepting, and always available. We have argued the appeal of these features may be best understood in terms of the demands and risks of human sociality. Under this framework, the lack of judgment, accessibility, and amiability of AI companions are not independent affordances, but rather are manifestations of the reduced relational complexity relative to human-human social interaction. SCAI is appealing precisely because it offers features of companionship ordinarily inseparable from this relational complexity.

We have identified six domains of relational complexity relevant to SCAI adoption: (i) relationship formation and dissolution; (ii) achieving mutual understanding; (iii) interpersonal motives and strategic uncertainty; (iv) social partner evaluation and self-presentation; (v) social embeddedness and information flow; and (vi) reciprocity and relationship maintenance. These features are interrelated, and uniting them under one framework helps to explain how structural features of AI companionship and integral features of human psychology work together to drive SCAI use.

While reduced relational complexity may be appealing due to its relative ease and lack of risk, adoption of social companions who lack the indelible features of human sociality may stifle the development and maintenance of skills necessary to succeed in the social world, while the deep levels of disclosure characteristic of SCAI use create privacy risks that may be obscured by the faux dyadic relationship. Although these potential outcomes remain to be tested, they follow from the same overarching feature that makes SCAI so compelling: reduced relational complexity offers some affordances of social interaction while minimizing the demands and risks of relating to another person.

Our approach is generative of multiple directions for future research. While we have pointed to suggestive evidence that indicates reduced relational complexity motivates user adoption of SCAI, much of this work relies on self-report data from qualitative studies. As such, experimental work is needed to test the effect of measurable reductions in features of relational complexity on user perceptions and related behavior. We have also outlined a series of predictions pertaining to the factors that drive users to turn to AI as a social companion. Additional research is needed to discern how the attenuation of specific sites of relational complexity affects SCAI engagement. This is particularly true for the costs of relationship formation, as well as for the cognitive burden of mentalizing and calculating social risk during interactions. At present, these two sites of complexity have received relatively little attention from scholars of SCAI. In addition, we suggest that research dealing with one aspect of relational complexity in human-AI interaction should consider how other related dimensions of relational complexity may affect user behavior and cognition. Finally, in light of the possible negative outcomes associated with extended SCAI use that we have described, future work should investigate the potentially heightened impacts of engaging with artificial social actors during childhood, a particularly critical period for social skill acquisition and relationship formation.

\printcredits

\section*{References}
\begin{description}
\item[] About Character.AI. (n.d.). Retrieved July 26, 2026, from \url{https://character.ai/about}
\item[] Ali, F., Zhang, Q., Tauni, M. Z., \& Shahzad, K. (2024). Social Chatbot: My Friend in My Distress. International Journal of Human--Computer Interaction, 40(7), 1702--1712. \url{https://doi.org/10.1080/10447318.2022.2150745}
\item[] Altman, I., \& Taylor, D. (1973). Social penetration: The development of interpersonal relationships. New York, Holt, Rinehart and Winston.
\item[] Anderson, K. S., O’Neil, E., White, B., Robb, M., Zimmermann, L., \& Mann, S. (2026). The Common Sense Media Census: AI Use by Tweens and Teens. Common Sense Media Youth AI Safety Institute.
\item[] Bariach, B., Schoenegger, P., Bhaskar, M., \& Suleyman, M. (2026). Seemingly conscious AI risks. AI and Ethics, 6(5), 455. \url{https://doi.org/10.1007/s43681-026-01294-x}
\item[] Baron-Cohen, S. (2000). Theory of mind and autism: A review. In Autism (Vol. 23, pp. 169--184). Academic Press. \url{https://doi.org/10.1016/S0074-7750(00)80010-5}
\item[] Bayor, L., Weinert, C., Maier, C., \& Weitzel, T. (2025). Social-Oriented Communication with AI Companions: Benefits, Costs, and Contextual Patterns. Business \& Information Systems Engineering, 67(5), 637--655. \url{https://doi.org/10.1007/s12599-025-00955-1}
\item[] Bland, A. R., Roiser, J. P., Mehta, M. A., Sahakian, B. J., Robbins, T. W., \& Elliott, R. (2022). The impact of COVID-19 social isolation on aspects of emotional and social cognition. Cognition \& Emotion, 36(1), 49--58. \url{https://doi.org/10.1080/02699931.2021.1892593}
\item[] Boiteau, T. W., Malone, P. S., Peters, S. A., \& Almor, A. (2014). Interference between conversation and a concurrent visuomotor task. Journal of Experimental Psychology: General, 143(1), 295--311. \url{https://doi.org/10.1037/a0031858}
\item[] Brandtzaeg, P. B., Skjuve, M., \& Følstad, A. (2022). My AI Friend: How Users of a Social Chatbot Understand Their Human--AI Friendship. Human Communication Research, 48(3), 404--429. \url{https://doi.org/10.1093/hcr/hqac008}
\item[] Brown, H., Lee, K., Mireshghllah, F., Shokri, R., \& Tramer, F. (2022). What Does it Mean for a Language Model to Preserve Privacy? 2022 ACM Conference on Fairness, Accountability, and Transparency (FAccT ’22). \url{https://doi.org/10.1145/3531146.3534642}
\item[] Brown-Schmidt, S., \& Hanna, J. E. (2011). Talking in another person’s shoes: Incremental perspective-taking in language processing. Dialogue \& Discourse, 2, 11--33. \url{https://doi.org/10.5087/dad.2011.102}
\item[] Brüne, M., \& Brüne-Cohrs, U. (2006). Theory of mind---Evolution, ontogeny, brain mechanisms and psychopathology. Neuroscience \& Biobehavioral Reviews, 30(4), 437--455. \url{https://doi.org/10.1016/j.neubiorev.2005.08.001}
\item[] Büttner, C. M., \& Lutz, S. (2025). When does waiting for a reply turn into ghosting? Individual, relational, and situational predictors of feeling ignored in online messaging. Computers in Human Behavior, 172, 108774. \url{https://doi.org/10.1016/j.chb.2025.108774}
\item[] Canary, D. J., \& Dainton, M. (2006). Maintaining Relationships. In The Cambridge handbook of personal relationships (pp. 727--743). Cambridge University Press. \url{https://doi.org/10.1017/CBO9780511606632.040}
\item[] Capuozzo, A., Rizzato, S., Grossi, G., \& Strappini, F. (2024). A Systematic Review on Social Cognition in ADHD: The Role of Language, Theory of Mind, and Executive Functions. Brain Sciences, 14(11), 1117. \url{https://doi.org/10.3390/brainsci14111117}
\item[] Carpenter, A., \& Greene, K. (2015). Social Penetration Theory. In The International Encyclopedia of Interpersonal Communication (pp. 1--4). John Wiley \& Sons, Ltd. \url{https://doi.org/10.1002/9781118540190.wbeic160}
\item[] Cerutti, J., Burt, K. B., Moeller, R. W., \& Seehuus, M. (2024). Declines in social--emotional skills in college students during the COVID-19 pandemic. Frontiers in Psychology, 15, 1392058. \url{https://doi.org/10.3389/fpsyg.2024.1392058}
\item[] Cheng, M., Lee, C., Khadpe, P., Yu, S., Han, D., \& Jurafsky, D. (2026). Sycophantic AI decreases prosocial intentions and promotes dependence. Science, 391(6792), eaec8352. \url{https://doi.org/10.1126/science.aec8352}
\item[] Chudek, M., Zhao, W., \& Henrich, J. (2013). Culture-gene coevolution, large-scale cooperation, and the shaping of human social psychology. In Cooperation and its evolution (pp. 425--457). The MIT Press. \url{https://doi.org/10.7551/mitpress/9033.001.0001}
\item[] Cinelli, M., De Francisci Morales, G., Galeazzi, A., Quattrociocchi, W., \& Starnini, M. (2021). The echo chamber effect on social media. Proceedings of the National Academy of Sciences, 118(9), e2023301118. \url{https://doi.org/10.1073/pnas.2023301118}
\item[] Clark, H. H., \& Brennan, S. E. (1991). Grounding in communication. In Perspectives on socially shared cognition (pp. 127--149). American Psychological Association. \url{https://doi.org/10.1037/10096-006}
\item[] Cohn, M., Pushkarna, M., Olanubi, G. O., Moran, J. M., Padgett, D., Mengesha, Z., \& Heldreth, C. (2024). Believing Anthropomorphism: Examining the Role of Anthropomorphic Cues on Trust in Large Language Models. Extended Abstracts of the CHI Conference on Human Factors in Computing Systems, CHI EA ’24, 1--15. \url{https://doi.org/10.1145/3613905.3650818}
\item[] Contreras-Huerta, L. S., Pisauro, M. A., \& Apps, M. A. J. (2020). Effort shapes social cognition and behaviour: A neuro-cognitive framework. Neuroscience and Biobehavioral Reviews, 118, 426--439. \url{https://doi.org/10.1016/j.neubiorev.2020.08.003}
\item[] Das, B. C., Amini, M. H., \& Wu, Y. (2025). Security and Privacy Challenges of Large Language Models: A Survey. ACM Comput. Surv., 57(6), 152:1-152:39. \url{https://doi.org/10.1145/3712001}
\item[] De Freitas, J. (2026). AI companions as hyper-attachment and caregiving targets. Current Opinion in Psychology, 73, 102393. \url{https://doi.org/10.1016/j.copsyc.2026.102393}
\item[] De Freitas, J., \& Cohen, I. G. (2025). Unregulated emotional risks of AI wellness apps. Nature Machine Intelligence, 7(6), 813--815. \url{https://doi.org/10.1038/s42256-025-01051-5}
\item[] De Freitas, J., Oğuz-Uğuralp, Z., \& Uğuralp, A. K. (2025). Emotional Manipulation by AI Companions (SSRN Scholarly Paper No. 5390377). Social Science Research Network. \url{https://doi.org/10.2139/ssrn.5390377}
\item[] Derks, D., Duin, D. van, Tims, M., \& Bakker, A. B. (2014). Smartphone use and work--home interference: The moderating role of social norms and employee work engagement. Journal of Occupational and Organizational Psychology, (88). \url{https://doi.org/10.1111/joop.12083}
\item[] DiFonzo, N., \& Bordia, P. (2007). Rumor, Gossip and Urban Legends. Diogenes, 54(1), 19--35. \url{https://doi.org/10.1177/0392192107073433}
\item[] Dores Cruz, T. D., Thielmann, I., Columbus, S., Molho, C., Wu, J., Righetti, F., de Vries, R. E., Koutsoumpis, A., van Lange, P. A. M., Beersma, B., \& Balliet, D. (2021). Gossip and reputation in everyday life. Philosophical Transactions of the Royal Society B: Biological Sciences, 376(1838), 20200301. \url{https://doi.org/10.1098/rstb.2020.0301}
\item[] Downey, G., \& Feldman, S. I. (1996). Implications of rejection sensitivity for intimate relationships. Journal of Personality and Social Psychology, 70(6), 1327--1343. \url{https://doi.org/10.1037//0022-3514.70.6.1327}
\item[] Downey, G., Freitas, A. L., Michaelis, B., \& Khouri, H. (1998). The self-fulfilling prophecy in close relationships: Rejection sensitivity and rejection by romantic partners. Journal of Personality and Social Psychology, 75(2), 545.
\item[] Dunbar, R. I. M. (1998). The social brain hypothesis. Evolutionary Anthropology: Issues, News, and Reviews, 6(5), 178--190.
\item[] Dunbar, R. I. M. (2004). Gossip in Evolutionary Perspective. Review of General Psychology, 8(2), 100--110. \url{https://doi.org/10.1037/1089-2680.8.2.100}
\item[] Eisenberger, N. I. (2012). The pain of social disconnection: Examining the shared neural underpinnings of physical and social pain. Nature Reviews Neuroscience, 13(6), 421--434. \url{https://doi.org/10.1038/nrn3231}
\item[] Eisenberger, N. I. (2015). Social pain and the brain: Controversies, questions, and where to go from here. Annual Review of Psychology, 66, 601--629. \url{https://doi.org/10.1146/annurev-psych-010213-115146}
\item[] Elder, C.-H., \& Beaver, D. (2022). “We’re running out of fuel!”: When does miscommunication go unrepaired? Intercultural Pragmatics, 19, 541--570. \url{https://doi.org/10.1515/ip-2022-5001}
\item[] Feinberg, M., Willer, R., \& Schultz, M. (2014). Gossip and Ostracism Promote Cooperation in Groups. Psychological Science, 25(3), 656--664. \url{https://doi.org/10.1177/0956797613510184}
\item[] Flathers, M., Roux, S., \& Torous, J. (2026). Beyond artificial intelligence psychosis: A functional typology of large language model-associated psychotic phenomena. The Lancet Digital Health, 8(4). \url{https://doi.org/10.1016/j.landig.2025.100974}
\item[] Fox, K. A., Nick, E., Nesi, J., Telzer, E. H., \& Prinstein, M. J. (2025). Why Haven’t You Texted Me Back? Adolescents’ Digital Entrapment, Friendship Conflict, and Perceived General Health. Journal of Clinical Child \& Adolescent Psychology, 54(5), 530--540. \url{https://doi.org/10.1080/15374416.2023.2261543}
\item[] Goddard, Q., Moton, N., Hudson, J., \& He, H. A. (2024). A Chatbot Won’t Judge Me: An Exploratory Study of Self-disclosing Chatbots in Introductory Computer Science Classes. Proceedings of the 26th Western Canadian Conference on Computing Education, WCCCE ’24, 1--7. \url{https://doi.org/10.1145/3660650.3660662}
\item[] Gottman, J. M., \& Driver, J. L. (2005). Dysfunctional Marital Conflict and Everyday Marital Interaction. Journal of Divorce \& Remarriage, 43(3--4), 63--77. https://doi.org/10.1300/J087v43n03\_04
\item[] Grice, H. P. (1975). Logic and Conversation. In D. Davidson (Ed.), The logic of grammar (pp. 64--75). Dickenson Pub. Co.
\item[] Guo, Y., Wang, R., Huang, Z., Jin, T., Yao, X., Feng, Y.-L., Zhang, W., Yao, Y., \& Mi, H. (2025). Exploring the Design of LLM-based Agent in Enhancing Self-disclosure Among the Older Adults. Proceedings of the 2025 CHI Conference on Human Factors in Computing Systems, CHI ’25, 1--17. \url{https://doi.org/10.1145/3706598.3713639}
\item[] Halford, G. S., Wilson, W. H., \& Phillips, S. (1998). Processing capacity defined by relational complexity: Implications for comparative, developmental, and cognitive psychology. Behavioral and Brain Sciences, 21(6), 803--831. \url{https://doi.org/10.1017/S0140525X98001769}
\item[] Hall, J. A. (2019). How many hours does it take to make a friend? Journal of Social and Personal Relationships, 36(4), 1278--1296. \url{https://doi.org/10.1177/0265407518761225}
\item[] Hall, J. A., \& Baym, N. K. (2012). Calling and texting (too much): Mobile maintenance expectations, (over)dependence, entrapment, and friendship satisfaction. New Media \& Society, 14(2), 316--331. \url{https://doi.org/10.1177/1461444811415047}
\item[] Hall, J. A., Horgan, T. G., \& Murphy, N. A. (2019). Nonverbal Communication. Annual Review of Psychology, 70, 271--294. \url{https://doi.org/10.1146/annurev-psych-010418-103145}
\item[] Hawkley, L. C., Cacioppo, J. T., \& Correll, J. (2013). Perceived Social Isolation within Personal and Evolutionary Timescales. In C. N. DeWall (Ed.), The Oxford Handbook of Social Exclusion (p. 0). Oxford University Press. \url{https://doi.org/10.1093/oxfordhb/9780195398700.013.0017}
\item[] Hays, R. B. (1984). The Development and Maintenance of Friendship. Journal of Social and Personal Relationships, 1(1), 75--98. https://doi.org/10.1177/0265407584011005
\item[] Haythornthwaite, C. (2005). Social networks and Internet connectivity effects. Information, Communication \& Society, 8(2), 125--147. \url{https://doi.org/10.1080/13691180500146185}
\item[] Heath, C. (1985). Talk and recipiency: Sequential organization in speech and body movement. In J. M. Atkinson (Ed.), Structures of Social Action (pp. 247--265). Cambridge University Press. \url{https://doi.org/10.1017/CBO9780511665868.017}
\item[] Heller, I., Bolliger, C., Holmer, P., Schladerer, S. P., \& Michel, G. (2026). The Role of Social Skills Training in the Treatment of Social Anxiety: A Systematic Review. Sage Open, 16(1), 21582440251409451. \url{https://doi.org/10.1177/21582440251409451}
\item[] Heng, S., \& Zhang, Z. (2025). Attachment Anxiety and Problematic Use of Conversational Artificial Intelligence: Mediation of Emotional Attachment and Moderation of Anthropomorphic Tendencies. Psychology Research and Behavior Management, 18, 1775--1785. \url{https://doi.org/10.2147/PRBM.S531805}
\item[] Henriksen, A., Asadi, R., Kulyk, O., Gerdes, A., \& Mayer, P. (2025). “I tell him everything that I do”: An investigation of privacy and safety implications of AI companion usage. 2025 European Symposium on Usable Security (EuroUSEC), 1--12. \url{https://doi.org/10.1109/EuroUSEC69254.2025.00011}
\item[] Hess, N. H., \& Hagen, E. H. (2023). The impact of gossip, reputation, and context on resource transfers among Aka hunter-gatherers, Ngandu horticulturalists, and MTurkers. Evolution and Human Behavior, Special Issue: Dispatches from the Field Part I, 44(5), 442--453. \url{https://doi.org/10.1016/j.evolhumbehav.2023.02.013}
\item[] Hill, K. (2025, June 13). They Asked an A.I. Chatbot Questions. The Answers Sent Them Spiraling. The New York Times. \url{https://www.nytimes.com/2025/06/13/technology/chatgpt-ai-chatbots-conspiracies.html}
\item[] Joshi, R., Adjagbodjou, A., \& Luria, M. (2026). Dark Patterns in AI Chatbots: A Taxonomy to Inform Better Design. Center for Democracy and Technology.
\item[] Kelley, H. H., Levinger, G. K., Berscheid, E., Levinger, G., McClintock, E., Peplau, L. A., Peterson, D. R., Christensen, A., Harvey, J. H., \& Huston, T. L. (1983). Close Relationships. W. H. Freeman and Company. \url{https://www.bookstores.com/books/close-relationships-harold-h-kelley/9780971242784}
\item[] Keysar, B., Lin, S., \& Barr, D. J. (2003). Limits on theory of mind use in adults. Cognition, 89(1), 25--41. \url{https://doi.org/10.1016/S0010-0277(03)00064-7}
\item[] Kilford, E. J., Garrett, E., \& Blakemore, S.-J. (2016). The development of social cognition in adolescence: An integrated perspective. Neuroscience and Biobehavioral Reviews, 70, 106--120. \url{https://doi.org/10.1016/j.neubiorev.2016.08.016}
\item[] Kim, T. W., Jiang, L., Duhachek, A., Lee, H., \& Garvey, A. (2022). Do You Mind if I Ask You a Personal Question? How AI Service Agents Alter Consumer Self-Disclosure. Journal of Service Research, 25(4), 649--666. \url{https://doi.org/10.1177/10946705221120232}
\item[] Kim, Y., \& Sundar, S. S. (2012). Anthropomorphism of computers: Is it mindful or mindless? Computers in Human Behavior, 28(1), 241--250. \url{https://doi.org/10.1016/j.chb.2011.09.006}
\item[] Kitzinger, C. (2012). Repair. In The Handbook of Conversation Analysis (pp. 229--256). John Wiley \& Sons, Ltd. \url{https://doi.org/10.1002/9781118325001.ch12}
\item[] Klein Ikkink, K., \& van Tilburg, T. (1999). Broken ties: Reciprocity and other factors affecting the termination of older adults’ relationships. Social Networks, 21(2), 131--146. \url{https://doi.org/10.1016/S0378-8733(99)00005-2}
\item[] Klitgaard, S. E., \& Fessler, D. M. T. (2026). An evolutionary approach to privacy and information-management psychology. Evolutionary Human Sciences, 8, e21. \url{https://doi.org/10.1017/ehs.2026.10051}
\item[] Kouroupa, A., Laws, K. R., Irvine, K., Mengoni, S. E., Baird, A., \& Sharma, S. (2022). The use of social robots with children and young people on the autism spectrum: A systematic review and meta-analysis. PLOS ONE, 17(6), e0269800. \url{https://doi.org/10.1371/journal.pone.0269800}
\item[] Kuzawa, C. W., Chugani, H. T., Grossman, L. I., Lipovich, L., Muzik, O., Hof, P. R., Wildman, D. E., Sherwood, C. C., Leonard, W. R., \& Lange, N. (2014). Metabolic costs and evolutionary implications of human brain development. Proceedings of the National Academy of Sciences, 111(36), 13010--13015. \url{https://doi.org/10.1073/pnas.1323099111}
\item[] Leary, M. R., \& Cottrell, C. A. (2013). Evolutionary Perspectives on Interpersonal Acceptance and Rejection. In C. N. DeWall (Ed.), The Oxford Handbook of Social Exclusion (p. 0). Oxford University Press. \url{https://doi.org/10.1093/oxfordhb/9780195398700.013.0002}
\item[] Lee, Y.-C., Yamashita, N., Huang, Y., \& Fu, W. (2020). “I Hear You, I Feel You”: Encouraging Deep Self-disclosure through a Chatbot. Proceedings of the 2020 CHI Conference on Human Factors in Computing Systems, CHI ’20, 1--12. \url{https://doi.org/10.1145/3313831.3376175}
\item[] Li, Z., Wu, Y., Li, Y., Gui, X., \& Luo, Y. (2026). Privacy Control in Conversational LLM Platforms: A Walkthrough Study. Proceedings of the 2026 CHI Conference on Human Factors in Computing Systems, CHI ’26, 1--26. \url{https://doi.org/10.1145/3772318.3791054}
\item[] Mai, L. M., Freudenthaler, R., Schneider, F. M., \& Vorderer, P. (2015). “I know you’ve seen it!” Individual and social factors for users’ chatting behavior on Facebook. Computers in Human Behavior, 49, 296--302. \url{https://doi.org/10.1016/j.chb.2015.01.074}
\item[] Maimann, K. (2025, September 17). AI-fuelled delusions are hurting Canadians. Here are some of their stories. CBC News. \url{https://www.cbc.ca/news/canada/ai-psychosis-canada-1.7631925}
\item[] Malfacini, K. (2025). The impacts of companion AI on human relationships: Risks, benefits, and design considerations. AI \& SOCIETY, 40(7), 5527--5540. \url{https://doi.org/10.1007/s00146-025-02318-6}
\item[] Malki, L. M. (2026). Towards Usable, Privacy Respecting Long-Term Memory for LLM-based Conversational Agents. Proceedings of the Extended Abstracts of the 2026 CHI Conference on Human Factors in Computing Systems, CHI EA ’26, 1--5. \url{https://doi.org/10.1145/3772363.3799198}
\item[] Maples, B., Pea, R. D., \& Markowitz, D. (2023). Learning from intelligent social agents as social and intellectual mirrors. In AI in learning: Designing the future (pp. 73--89). Springer Nature Switzerland AG. \url{https://doi.org/10.1007/978-3-031-09687-7\_5}
\item[] Marriott, H. R., \& Pitardi, V. (2024). One is the loneliest number… Two can be as bad as one. The influence of AI Friendship Apps on users’ well-being and addiction. Psychology \& Marketing, 41(1), 86--101. \url{https://doi.org/10.1002/mar.21899}
\item[] Mildner, J. N., \& Tamir, D. I. (2021). The people around you are inside your head: Social context shapes spontaneous thought. Journal of Experimental Psychology. General, 150(11), 2375--2386. \url{https://doi.org/10.1037/xge0001057}
\item[] Modgil, S., Singh, R. K., Gupta, S., \& Dennehy, D. (2024). A Confirmation Bias View on Social Media Induced Polarisation During Covid-19. Information Systems Frontiers, 26(2), 417--441. \url{https://doi.org/10.1007/s10796-021-10222-9}
\item[] Morrison, A. S., \& Heimberg, R. G. (2013). Social Anxiety and Social Anxiety Disorder. Annual Review of Clinical Psychology, 9, 249--274. \url{https://doi.org/10.1146/annurev-clinpsy-050212-185631}
\item[] Nass, C., \& Moon, Y. (2000). Machines and Mindlessness: Social Responses to Computers. Journal of Social Issues, 56(1), 81--103. \url{https://doi.org/10.1111/0022-4537.00153}
\item[] Nass, C., Steuer, J., \& Tauber, E. R. (1994). Computers are social actors. Proceedings of the SIGCHI Conference on Human Factors in Computing Systems, CHI ’94, 72--78. \url{https://doi.org/10.1145/191666.191703}
\item[] Navarro, E., Macnamara, B. N., Glucksberg, S., \& Conway, A. R. A. (2020). What influences successful communication? An examination of cognitive load and individual differences. Discourse Processes, 57(10), 880--899. \url{https://doi.org/10.1080/0163853X.2020.1829936}
\item[] Neff, G., \& Nagy, P. (2025). The quasi-domestication of social chatbots: The case of Replika. New Media \& Society, 27(10), 5508--5524. \url{https://doi.org/10.1177/14614448251359218}
\item[] Neff, L. A., \& Broady, E. F. (2011). Stress resilience in early marriage: Can practice make perfect? Journal of Personality and Social Psychology, 101(5), 1050--1067. \url{https://doi.org/10.1037/a0023809}
\item[] Nikolajczyk, M., Maier, E., Prawitz, J., \& Dogruel, L. (2026). Beyond Human Bonds: Understanding Relationship Development with Humanized AI Companions. Human-Machine Communication, 13(1). \url{https://doi.org/10.30658/hmc.13.7}
\item[] Noorbehbahani, F., \& Oyibo, K. (2026). AI-overdependence and human cognitive decline: Hazards, evidence, and mitigation strategies. Computers in Human Behavior Reports, 22, 101102. \url{https://doi.org/10.1016/j.chbr.2026.101102}
\item[] OpenAI. (2025, October 27). Strengthening ChatGPT’s responses in sensitive conversations. OpenAI. \url{https://openai.com/index/strengthening-chatgpt-responses-in-sensitive-conversations/}
\item[] OpenAI. (2026, July 22). How people are using ChatGPT. OpenAI. \url{https://openai.com/index/how-people-are-using-chatgpt/}
\item[] Panfili, L., Duman, S., Nave, A., Ridgeway, K. P., Eversole, N., \& Sarikaya, R. (2021). Human-AI interactions through a Gricean lens. Proceedings of the Linguistic Society of America, 6(1), Article 1. \url{https://doi.org/10.3765/plsa.v6i1.4971}
\item[] Papneja, H., \& Yadav, N. (2025). Self-disclosure to conversational AI: A literature review, emergent framework, and directions for future research. Personal and Ubiquitous Computing, 29(2), 119--151. \url{https://doi.org/10.1007/s00779-024-01823-7}
\item[] Rendle-Short, J. (2015). Dispreferred responses when texting: Delaying that ‘no’ response. Discourse \& Communication, 9(6), 643--661. \url{https://doi.org/10.1177/1750481315600309}
\item[] Replika is the AI Friend to Do Life With. (n.d.). Replika. Retrieved July 26, 2026, from \url{https://replika.com}
\item[] Ríos, J. C., \& Fernández, L. (2025). Hidden Grief, Visible Grades: Components of Academic Functioning after Friendship Loss in Adolescents. Applied Family Therapy Journal (AFTJ), 1--10.
\item[] Risko, E. F., \& Gilbert, S. J. (2016). Cognitive Offloading. Trends in Cognitive Sciences, 20(9), 676--688. \url{https://doi.org/10.1016/j.tics.2016.07.002}
\item[] Romero-Canyas, R., Downey, G., Berenson, K., Ayduk, O., \& Kang, N. J. (2010). Rejection Sensitivity and the Rejection--Hostility Link in Romantic Relationships. Journal of Personality, 78(1), 119--148. \url{https://doi.org/10.1111/j.1467-6494.2009.00611.x}
\item[] Saygin, A. P., \& Cicekli, I. (2002). Pragmatics in human-computer conversations. Journal of Pragmatics, 34(3), 227--258. \url{https://doi.org/10.1016/S0378-2166(02)80001-7}
\item[] Scassellati, B., Boccanfuso, L., Huang, C.-M., Mademtzi, M., Qin, M., Salomons, N., Ventola, P., \& Shic, F. (2018). Improving social skills in children with ASD using a long-term, in-home social robot. Science Robotics, 3(21), eaat7544. \url{https://doi.org/10.1126/scirobotics.aat7544}
\item[] Schacter, H. L., Lessard, L. M., \& Juvonen, J. (2019). Peer rejection as a precursor of romantic dysfunction in adolescence: Can friendships protect? Journal of Adolescence, 77(1), 70--80. \url{https://doi.org/10.1016/j.adolescence.2019.10.004}
\item[] Schegloff, E. (2007). Sequence Organization in Interaction: A Primer in Conversation Analysis. Sequence Organization in Interaction: A Primer in Conversation Analysis I, 1, 1--300. \url{https://doi.org/10.1017/CBO9780511791208}
\item[] Skjuve, M., Brandtzaeg, P. B., \& Følstad, A. (2024). Why do people use ChatGPT? Exploring user motivations for generative conversational AI. First Monday. \url{https://doi.org/10.5210/fm.v29i1.13541}
\item[] Skjuve, M., Følstad, A., Fostervold, K. I., \& Brandtzaeg, P. (2021). My Chatbot Companion---A Study of Human-Chatbot Relationships. International Journal of Human-Computer Studies. \url{https://doi.org/10.1016/j.ijhcs.2021.102601}
\item[] Sperber, D., \& Wilson, D. (1995). Relevance: Communication and cognition (2nd ed., pp. viii, 326). Blackwell Publishing.
\item[] Spitale, M. (2019). Training self-sufficiency and social skills with embodied conversational agent for children with autism. Companion Proceedings of the 24th International Conference on Intelligent User Interfaces, IUI ’19 Companion, 151--152. \url{https://doi.org/10.1145/3308557.3308722}
\item[] Staab, R., Vero, M., Balunovic, M., \& Vechev, M. (2024). Beyond Memorization: Violating Privacy via Inference with Large Language Models. International Conference on Learning Representations, 2024, 33832--33878.
\item[] Sun, X., Wang, Y., \& McDaniel, B. T. (2026). AI companions and adolescent social relationships: Benefits, risks, and bidirectional influences. Child Development Perspectives, aadaf009. \url{https://doi.org/10.1093/cdpers/aadaf009}
\item[] Tangermann, V. (2025, June 13). Man Killed by Police After Spiraling Into ChatGPT-Driven Psychosis. Futurism. \url{https://futurism.com/man-killed-police-chatgpt}
\item[] Tomasello, M. (2008). Origins of Human Communication. The MIT Press. \url{https://doi.org/10.7551/mitpress/7551.001.0001}
\item[] Turkle, S. (2011). Alone Together: Why We Expect More from Technology and Less from Each Other. Basic Books.
\item[] Turkle, S. (2026). Artificial Intimacy: Who We Become When We Talk to Machines. Little, Brown and Company.
\item[] Twenge, J. M., \& Spitzberg, B. H. (2020). Declines in non-digital social interaction among Americans, 2003--2017. Journal of Applied Social Psychology, 50(6), 363--367. \url{https://doi.org/10.1111/jasp.12665}
\item[] Twenge, J. M., Spitzberg, B. H., \& Campbell, W. K. (2019). Less in-person social interaction with peers among U.S. adolescents in the 21st century and links to loneliness. Journal of Social and Personal Relationships. \url{https://doi.org/10.1177/0265407519836170}
\item[] Vohs, K. D., Baumeister, R. F., \& Ciarocco, N. J. (2005). Self-regulation and self-presentation: Regulatory resource depletion impairs impression management and effortful self-presentation depletes regulatory resources. Journal of Personality and Social Psychology, 88(4), 632--657. \url{https://doi.org/10.1037/0022-3514.88.4.632}
\item[] Wahlberg, T., \& Magliano, J. P. (2004). The Ability of High Function Individuals With Autism to Comprehend Written Discourse. Discourse Processes, 38(1), 119--144. \url{https://doi.org/10.1207/s15326950dp3801\_5}
\item[] Williams, K. D. (2007). Ostracism. Annual Review of Psychology, 58(Volume 58, 2007), 425--452. \url{https://doi.org/10.1146/annurev.psych.58.110405.085641}
\item[] Williams, K. D., \& Zadro, L. (2005). Ostracism: The indiscriminate early detection system. In The social outcast: Ostracism, social exclusion, rejection, and bullying (pp. 19--34). Psychology Press.
\item[] Wilson, A. C., \& Bishop, D. V. M. (2021). “Second guessing yourself all the time about what they really mean…”: Cognitive differences between autistic and non-autistic adults in understanding implied meaning. Autism Research, 14(1), 93--101. \url{https://doi.org/10.1002/aur.2345}
\item[] Wood, N. R., Wicks, S. G., Beam, A. J., Mudryk, E. P., Bray, E., \& Hales, A. H. (2026). What Could Have Been: Predicted and Actual Exclusion by Potential Romantic Partners and Platonic Friends. European Journal of Social Psychology, 56(4), 674--690. \url{https://doi.org/10.1002/ejsp.70066}
\item[] Wu, H., Fung, B. J., \& Mobbs, D. (2022). Mentalizing During Social Interaction: The Development and Validation of the Interactive Mentalizing Questionnaire. Frontiers in Psychology, 12. \url{https://doi.org/10.3389/fpsyg.2021.791835}
\item[] Wu, H., Liu, X., Hagan, C. C., \& Mobbs, D. (2020). Mentalizing during social InterAction: A four component model. Cortex; a Journal Devoted to the Study of the Nervous System and Behavior, 126, 242--252. \url{https://doi.org/10.1016/j.cortex.2019.12.031}
\item[] Wu, J., Balliet, D., \& Van Lange, P. A. M. (2016). Reputation management: Why and how gossip enhances generosity. Evolution and Human Behavior, 37(3), 193--201. \url{https://doi.org/10.1016/j.evolhumbehav.2015.11.001}
\item[] Xie, T., \& Pentina, I. (2022). Attachment Theory as a Framework to Understand Relationships with Social Chatbots: A Case Study of Replika. \url{https://doi.org/10.24251/HICSS.2022.258}
\item[] Xue, M., \& Silk, J. B. (2012). The role of tracking and tolerance in relationship among friends. Evolution and Human Behavior, 33(1), 17--25. \url{https://doi.org/10.1016/j.evolhumbehav.2011.04.004}
\item[] Yoon, S., \& Kang, Y. (2026). Are we close enough? Impact of stress management chatbot’s self-disclosure and relationship types on user’s self-disclosure, trust, intimacy, and anonymity. International Journal of Human-Computer Studies, 209, 103738. \url{https://doi.org/10.1016/j.ijhcs.2026.103738}
\item[] Zhang, P., Yuan, Y., Wang, M., Ding, L., Zhang, J., \& Qiu, S. (2025). The relationship between social exclusion and social anxiety among Chinese college students: A chain mediation role of fear of negative evaluation and rejection sensitivity. Current Psychology, 44(8), 7181--7192. \url{https://doi.org/10.1007/s12144-025-07689-z}
\item[] Zhang, Y., Zhao, D., Hancock, J. T., Kraut, R., \& Yang, D. (2026). Interaction with AI companions and psychological well-being. Nature Human Behaviour, 1--15. \url{https://doi.org/10.1038/s41562-026-02516-2}
\item[] Zimmer-Gembeck, M. J., Nesdale, D., McGregor, L., Mastro, S., Goodwin, B., \& Downey, G. (2013). Comparing reports of peer rejection: Associations with rejection sensitivity, victimization, aggression, and friendship. Journal of Adolescence, 36(6), 1237--1246. \url{https://doi.org/10.1016/j.adolescence.2013.10.002}
\end{description}
\end{document}